\documentclass{emulateapj}
\usepackage{natbib}

\usepackage{epsfig}

\newcommand{\be}{\begin{equation}}
\newcommand{\ee}{\end{equation}}
\newcommand{\ba}{\begin{eqnarray}}
\newcommand{\ea}{\end{eqnarray}}

\begin{document}

\title{Supernovae, Lensed CMB and  Dark Energy}

\author{Wayne Hu$^{1,2}$, Dragan Huterer$^{1,2}$, and Kendrick M. Smith$^{1,3}$  }
\affil{ $^{1}$Kavli Institute for Cosmological Physics, 
Enrico Fermi Institute, University of Chicago, Chicago, IL 60637\\
$^{2}$Department of Astronomy and Astrophysics, University of Chicago, Chicago, IL 60637\\
$^{3}$Department of Physics,  University of Chicago, Chicago, IL 60637}

\begin{abstract}
Supernova distance and primary CMB anisotropy measurements provide
powerful probes of the dark energy evolution in a  flat universe but degrade
substantially once curvature is marginalized.  We show that lensed CMB polarization
power spectrum measurements, accessible to next generation ground based surveys
such as SPTpol or QUIET, can remove the curvature degeneracy at a level
sufficient for the SNAP and Planck surveys and allow a measurement of
$\sigma(w_p)=0.03$, $\sigma(w_a)=0.3$ jointly with $\sigma(\Omega_K)=0.0035$.
This expectation assumes that the sum of neutrino masses is
independently known to better than $0.1$ eV.  This assumption is valid
 if the lightest neutrino is assumed to have negligible mass in a
normal neutrino mass hierarchy and is potentially testable 
with upcoming direct laboratory measurements. 
\end{abstract}

\section{Introduction}
\label{sec:introduction}

Currently, observations of the expansion history of the universe are remarkably
consistent with cosmic acceleration driven by a cosmological constant in a spatially
flat universe. 
When testing this hypothesis, one typically looks for possible evidence of
spatial curvature in the absence of dark energy evolution {\it or} evolution in the
absence of spatial curvature.   It is of course possible that spatial curvature
and dark energy evolution conspire to mimic a cosmological constant in 
a flat universe.  Nonetheless while the data remain consistent with the simpler
hypothesis, this approach is justified.

More
worrying is the possibility that as measurements improve, 
we find evidence for non-standard dark energy in a flat universe --- a
dark energy equation of state $w \ne -1$. Should we then 
believe that the universe is flat and dark energy varying in time, or
that it has a small curvature and the dark energy is simply the cosmological
constant? 
While the standard inflationary
theory predicts that the curvature of our Hubble volume is below measurable
limits ($\Omega_K\lesssim 10^{-4}$),  models that allow a detectable spatial
curvature do exist and are arguably
on sounder footing than dynamical dark energy models.

Ideally of course, we would like to measure both $\Omega_K$ and $w(z)$ but this
is difficult because of degeneracies. 
Moderately good constraints are obtained once
type Ia supernova (SNe) data and cosmic microwave background (CMB)
data are combined with high precision Hubble constant measurements \citep{Hu04b,linder},
weak
gravitational lensing \citep{knox,bernstein}, baryon oscillations
\citep{knox_song_zhan,ichikawa} or cluster abundances. However, these
techniques are
subject to a vast array of systematic uncertainties that have to be accounted
for carefully.   For example,  
weak gravitational lensing  requires
modeling of the fully nonlinear power on scales of 0.1-100
Mpc to the accuracy of a few percent \citep[e.g.,][]{hut_takada}.

Only two
methods -- SNe Ia and CMB -- have proven so far to be both powerful and 
robust probes of
cosmology. Here we show that the information required to break the degeneracy between
curvature and dark energy to a level sufficient for future SNe missions such
as SNAP \citep{SNAP}
 lies within
the reach of next generation ground-based CMB polarization
power spectrum measurements. 
This information comes from weak gravitational lensing of the CMB in the
linear regime at redshifts $z \sim  1-3$ (see \cite{lewis_review} for a recent review).
We employ a recently developed framework for CMB lensing power spectrum
observables that includes
the non-Gaussian nature of the lensing signal \citep{SmiHuKap06}.   This method is
ideally suited for investigating the complementarity between
different cosmological probes in a wide range of dark energy models.

\section{Methodology}
\label{sec:methodology}

To describe the information content of the various cosmological probes, 
we model the observables and employ the usual Fisher approach. 
For SNe Ia, we model the magnitudes $m_i$ of the SNe as 
\begin{equation}
m_i = 5 \log H_0 d_L(z_i) + \mathcal{M} + \epsilon_i \,,
\end{equation}
where $i$ runs through the observed SNe.  Here 
the luminosity distance is given by 
\begin{equation}
d_L(z) = (1+z) { 1\over \sqrt{\Omega_K H_0^2}} \sinh\left( \sqrt{\Omega_K H_0^2} D \right)\,,
\end{equation}
where $\Omega_K$ is the curvature in units of the critical energy density, $H_0$ is the Hubble constant, $D(z)=\int dz/H(z)$ is the comoving radial distance, $\mathcal{M}\equiv
M-5\log H_0+25$ is a nuisance parameter involving the unknown absolute
magnitude of the supernova $M$.  
The noise term
$\epsilon_i$ represents both statistical errors and possible systematic errors
that do not necessarily decrease with the number of observed supernovae.

\begin{figure}[ht]
\centerline{\psfig{file=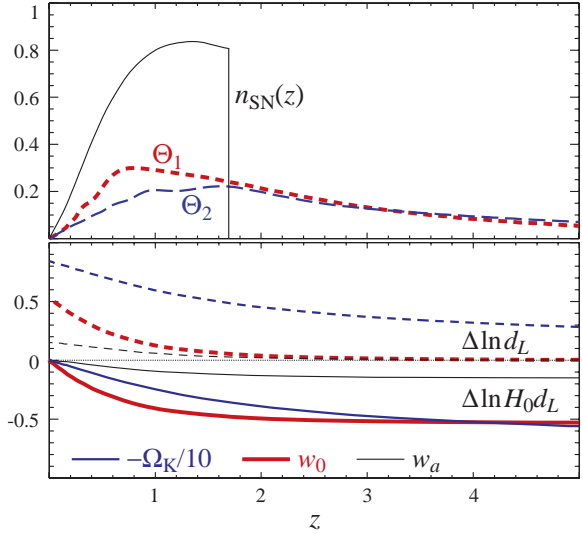, width=3.0in}}
  \caption{Top panel: redshift distribution of SNAP supernovae $n_{\rm SN}$ and weights of the lensing
  observables \{$\Theta_1,\Theta_2$\} normalized to integrate
  to unity.  Bottom panel: derivatives of the luminosity
  distance $d_L$ and relative luminosity distance $H_0 d_L$ with respect to
  curvature $\Omega_K$ and the dark energy parameters $w_0$ and $w_a$.  Here
  $\Omega_{\rm DE}$ is adjusted to keep the angular diameter distance to
  recombination fixed in each case.}
  \label{fig:redshift}
\end{figure}

We assume a
survey similar to the planned SNAP mission \citep{SNAP_WP} with $2800$
SNe
distributed in redshift out to $z=1.7$ given by
 \citet{SNAP} (middle curve of their Fig. 9, reproduced here in
Fig.~\ref{fig:redshift}).  We combine the SNAP dataset with 
300 local supernovae uniformly distributed in the $z=0.03-0.08$ range.

Following \cite{SNAP_WP}, we model the error 
as a sum of the statistical error and an irreducible, but unbiased,
systematic error. The latter imposes a floor on the errors at a given redshift that
is uncorrelated across broad redshift ranges.  
Given a binning of SNe into some arbitrary bins in $z$ denoted as $\Delta z_I$, we
assume that 
\begin{equation}
\sum_{i \in I,j \in J}{ {\langle( m_i -\bar m_i) (m_j -\bar m_j) \rangle }\over N_{I}N_{J}}= 
\delta_{IJ}\left ( { \sigma_m^2 \over N_I} +  \sigma_{\rm sys}^2\right )\,,
\end{equation}
where $N_I$ is the number of SNe in $\Delta z_I$.  Following
\cite{TegEisHuKro98}, we can replace the sum over discrete SNe with an integral
over the redshift distribution, $N_I = n_{\rm SN}(z) \Delta z_I$ and construct
the Fisher matrix for a parameter set $p_\mu$ as
\begin{equation}
F_{\mu \nu}^{\rm SNAP} = \int d z n_{\rm SN}(z)  {1 \over \sigma_\epsilon^2(z)} 
{\partial \bar m(z) \over \partial p_\mu} {\partial \bar m (z) \over \partial p_\nu} \,,
\end{equation}
where 
\begin{equation}
\sigma_\epsilon^2 = \sigma_m^2  + \sigma_{\rm sys}^2 n_{\rm SN}(z)\Delta z  \,.
\end{equation}
For SNAP, we take $\sigma_m = 0.15$ and $\sigma_{\rm sys} = 0.02\,(1+z)/2.7$, 
$\Delta z=0.1$.
When constructing the Fisher matrix in cosmological parameters we marginalize $\mathcal{M}$.

\begin{figure}[ht]
\centerline{\psfig{file=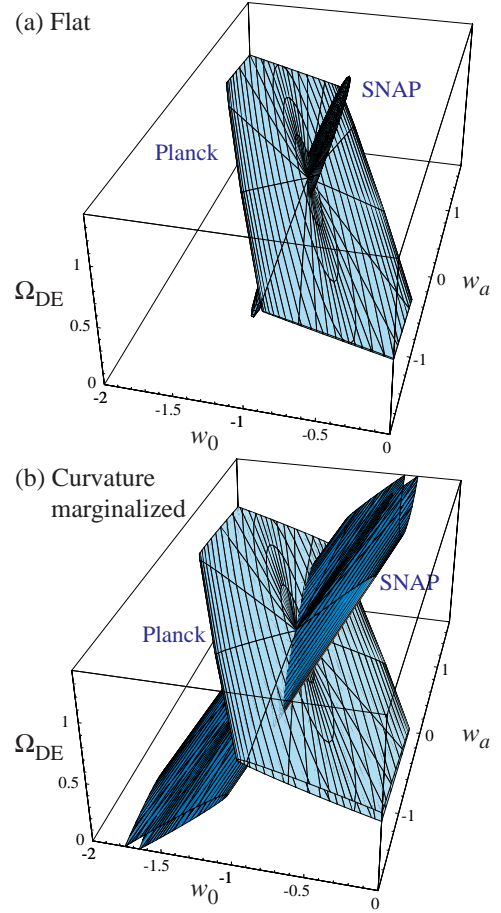, width=2.5in}}
  \caption{Constraints on dark energy parameters $\Omega_{\rm DE}$, $w_0$ and $w_a$,
shown for supernovae (SNAP) and unlensed CMB (Planck) separately:
(a) constraints assuming a flat universe; (b) weakened constraints
from curvature marginalization.}
  \label{fig:3dplot}
\end{figure}

For the CMB, we use the information coming from recombination by constructing the
Fisher matrix of the unlensed CMB out to multipole $\ell = 2000$ in the usual way
\citep{ZalSpeSel97}.
We assume the Planck survey with 80\% usable sky and 
3 usable channels for cosmology: FHWM $5.0'$ 
with temperature noise $\Delta_T = 51\mu$K$'$ and polarization noise
$\Delta_P = 135\mu$K$'$ ; $7.1'$ with $\Delta_T=43\mu$K$'$, $\Delta_P=78\mu$K$'$;
$9.2'$ with $\Delta_T = 51\mu$K$'$, $\Delta_P =\infty$.
We will call this Fisher matrix $F_{\mu\nu}^{\rm Planck}$.

For the additional information from lensing, we use the lensing observables
framework \citep[see][for details]{SmiHuKap06}.  Constraints below so derived are an
excellent match to those obtained through the full non-Gaussian band power covariance
matrix.   The two lensing observables
$\Theta_1$ and $\Theta_2$ are determined from temperature/$E$-polarization
 and $B$-polarization respectively.  They are 
 associated with the amplitude of the convergence
power spectrum in broad bands around $\ell_1 \sim 100$ and $\ell_2 \sim 500$.  
This amplitude in turn has a redshift
sensitivity plotted in Fig.~\ref{fig:redshift}.  Note that this sensitivity
extends to $z\gg 1$ and is reason that CMB lensing has higher sensitivity to
curvature than dark energy.  The CMB lensing Fisher matrix is given by
\begin{equation}
F_{\mu\nu}^{\rm CMBlens} =\sum_{i=1,2} {\partial\Theta_i \over\partial p_\mu}
{1 \over  \sigma^{2}_{\Theta_i}} {\partial\Theta_i \over\partial p_\nu} \,.
\label{eqn:FCMBlens}
\end{equation}

For the errors on the observables, 
we will assume a deep CMB survey that is comparable to the proposed SPTpol
survey.  Specifically we take a deep temperature
survey on
4000 deg$^2$ and $\Delta_T = $ 11.5$\mu $K$'$ and a deep polarization 
survey on 
625 deg$^{2}$ with $\Delta_P= \sqrt{2} \Delta_T = 4\mu$K$'$.  We take a FWHM beam of
1'.  
With these specifications combined with sensitivity to $\Theta_1$ from Planck,
 the two observables can be measured with an accuracy of
$\sigma_{\Theta_1} =0.041 $ and $\sigma_{\Theta_2}=0.032$.  For reference, the latter
represents a $\sim 3\%$ measurement of the overall power in lensing $B$-modes and
dominates the overall constraints.  Moreover the deep temperature survey provides
little weight in the $\Theta_1$ constraint itself
and would mainly serve as an internal cross check for foregrounds, systematics
and other secondaries.  
Likewise,
other planned surveys such as QUIET will have comparable precision in $\Theta_2$ with
very different frequency bands.

Finally, we  sum the Fisher matrices as usual
\begin{equation}
F_{\mu\nu} = F_{\mu \nu}^{\rm SNAP} + F_{\mu \nu}^{\rm Planck} + F_{\mu\nu}^{\rm CMBlens}
\end{equation}
and approximate the joint parameter covariance matrix as $C_{\mu\nu}= ({\bf F})^{-1}_{\mu\nu}$.

\begin{figure}[t]
\centerline{\psfig{file=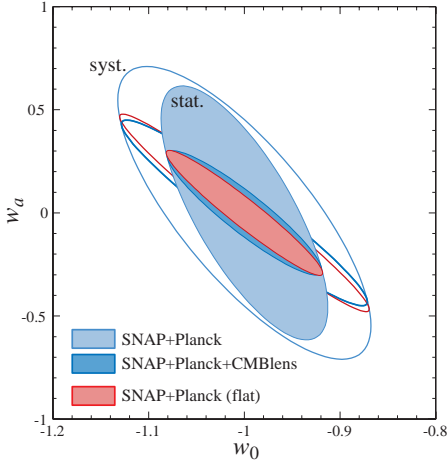, width=2.5in}}
  \caption{68\% CL region in $w_0-w_a$ with and without CMB lensing (CMBlens)
  information. Filled ellipses:
  SNAP statistical errors.  Open ellipses: SNAP systematic and statistical
  errors. Inner ellipse in each set: Planck+SNAP errors alone in a flat
  universe. 
  }
  \label{fig:improvement}
\end{figure}

\section{Forecasts with Curvature}

It is well known that CMB information from recombination allows
SNe to determine the dark energy equation of state parameters
\begin{equation}
w(a) = w_{0} + (1-a)w_{a}\,,
\end{equation}
in a flat universe.  In the 3-dimensional space \{$\Omega_{\rm DE}$(=0.76), 
$w_{0}(=-1)$, $w_{a}(=0)$\}, Planck CMB measurements limit the allowed region
to a 2-dimensional surface or plane
in the Fisher approximation (see Fig.~\ref{fig:3dplot}).  Values in parentheses represent those of
 the fiducial model.  Here we have marginalized
the baryon density $\Omega_{b}h^{2}(=0.022)$, cold dark matter density 
$\Omega_{c}h^{2}(=0.106)$,
tilt $n_{s}(=0.958)$, initial amplitude of curvature
fluctuations $\delta_{\zeta}(=4.52 \times 10^{-5})$ at $k=0.05/{\rm Mpc}$, 
and reionization optical depth $\tau(=0.92)$.

SNAP supernovae measurements constrain a flat tube in this space that is
nearly orthogonal to the Planck surface.  Note that the pre-marginalization of
any one of the three parameters before combining does not bring out this
complementarity or the quality of the two surveys in confining the volume in
the allowed dark energy space.

The marginalization of spatial curvature can be visualized as the superposition
of independent shifts in the Planck plane and SNAP tube.  Given that even the
unlensed CMB has distance-independent, albeit weak, curvature information from both the
ISW effect and the acoustic peaks
\citep[see][Fig.~11]{HuWhi96b}, the Planck plane only widens marginally (see
Fig.~\ref{fig:3dplot}).  On the other hand the SNAP tube widens substantially.
The net effect on the joint constraints in the $w_{0}-w_{a}$ plane marginalized
over \{$\Omega_{\rm DE}$, $\Omega_{K}$\}
is shown in 
Fig.~\ref{fig:improvement}.
It represents a factor of 4.8 in the
68\% CL area  \citep{hut_turner} as measured by
$A_{w}=\sigma(w_p)\sigma(w_{a})$.
Here $w_p$ is the equation of state at the best constrained or pivot redshift
and its errors are equal to those of $w_0$ at fixed $w_a$
\citep{HuJai03}.

This degeneracy is also illustrated in Fig.~\ref{fig:redshift} (bottom panel).  Here
the fractional deviations in the SNe observable $H_{0}d_{L}$ from the fiducial model 
are shown as parameter derivatives at a fixed distance to recombination. 
Without spatial curvature, $w_{0}$ and $w_{a}$ 
make distinguishable changes in the relative distance at $z<2$.  With spatial curvature,
the effects become largely degenerate.

The effect of spatial curvature on observables
persists to high redshift $z\gg 1$ whereas that of the dark energy parameters
flatten and depend only on $H_{0}$, the difference between relative and
absolute distances.  This degeneracy may therefore be broken either by high precision
Hubble constant  \citep{Hu04b,linder} or high-$z$ distance
 measurements \citep{knox,bernstein}.  

\begin{figure}[t]
\centerline{\psfig{file=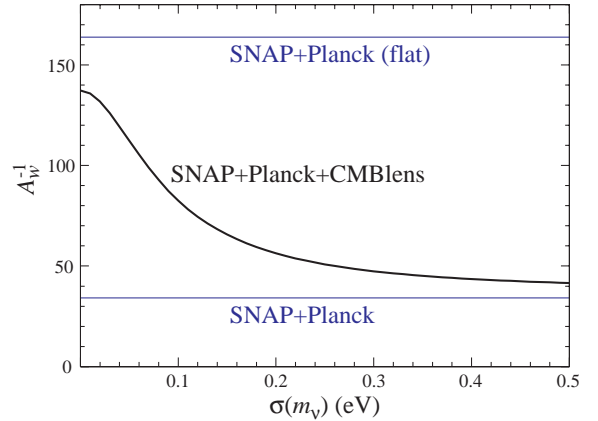, width=3.0in}}
  \caption{Improvement in the area statistic $A_{w}$
   in the $w_0-w_a$ plane of  as a function of the prior
  on the sum of the neutrino masses.  The top line represents the SNAP+Planck
  constraint alone in  a flat universe, the bottom represents the degradation once
  curvature is marginalized.   CMB lensing can recover much of this information if
  the sum of the neutrino masses is known to $\sigma(m_\nu) < 0.1$eV.}
  \label{fig:neutrinos}
\end{figure}

CMB lensing supplies the latter kind of information.  Around the fiducial model
the sensitivity of the lensing observables 
to cosmological parameters used in
Eqn.~(\ref{eqn:FCMBlens}) is
\begin{eqnarray}
\Delta\Theta_1 &\approx& 
-1.01 \Omega_{\rm DE}
-0.399\Delta w_0
-0.146\Delta w_a
-5.17\Delta\Omega_K \nonumber\\
&&
+12.3 \Delta\Omega_c h^2
+2\Delta\ln\delta_{\zeta}
-0.33{\Delta m_{\nu} \over 1{\rm eV}} \,, \nonumber\\
\Delta\Theta_2 &\approx& 
-1.27 \Omega_{\rm DE}
-0.446\Delta w_0
-0.154\Delta w_a
-5.30\Delta\Omega_K \nonumber\\
&&
+18.8 \Delta\Omega_c h^2
+2.09\Delta\ln\delta_{\zeta}
-0.45{\Delta m_{\nu} \over 1{\rm eV}} \,.
\end{eqnarray}
This lensing sensitivity depends both on parameters that control distances and
the matter power spectrum at the redshift range shown in
Fig.~\ref{fig:redshift}.
We have here used the fact
 that other parameters in the latter class such as $n_{s}$ are sufficiently well determined by
 Planck.   The sum of the neutrino masses $m_\nu$ is however not well determined and
 changes both the shape and growth rate of the matter power spectrum.
 The distance parameters also change the growth rate and the curvature sensitivity is
 in fact enhanced by this effect.

First let us consider the impact of CMB lensing constraints assuming that
the sum of the neutrino masses $m_{\nu}(=0.06$eV) is fixed.  This is
a good assumption if the lightest neutrino has a mass $<0.01$eV and
a normal mass hierarchy due to the measurement of the solar and
atmospheric neutrino mass squared differences.  The same assumption
in an inverted hierarchy would also be sufficient in that it only adds
a second discrete possibility.

In this fixed neutrino case, the addition of the lensing constraint nearly fully restores
the ability of SNAP and Planck to measure the dark energy (see Fig.~\ref{fig:improvement}).
It allows a measurement to $\sigma(w_{a}) = 0.30$ and 
$A_{w}^{-1}=137$.  This restoration of sensitivity occurs even
if SNAP is limited by only statistical errors such that $\sigma(w_{a}) = 0.19$ and 
$A_{w}^{-1}=241$. 

Figure \ref{fig:neutrinos} shows how $A_{w}^{-1}$ depends on 
prior knowledge of the sum of the neutrino masses in the case that the lightest
neutrino does not have negligible mass.  In this case all three neutrinos could
have degenerate masses.   
External constraints on the sum of neutrino masses begin to help at the $0.2$eV 
level and would be fully sufficient at a few $10^{-2}$eV.
 For example the KATRIN experiment
is expected to reach $\sigma(m_{\nu_e}^2) = (0.16{\rm eV})^2$ from tritium $\beta$ decay
\citep{KATRIN}.  Such a measurement would test the degenerate mass scenario.

As an aside it is interesting to note that even with spatial curvature and dark
energy the combination of data sets would allow a measurement of $\sigma(m_{\nu})=0.24$ eV.
With curvature fixed, $\sigma(m_{\nu})=0.14$ eV.   Hubble
constant measurements with $1-7\%$ precision would provide neutrino measurements
that interpolate between these two limits by fixing the spatial curvature.

\section{Discussion}

Constraints on the temporal evolution of dark energy benefit particularly
strongly from the addition of CMB lensing information to that of SNe and the primary CMB at
recombination. The three methods probe very
different epochs: SNe are sensitive to distances at $z\lesssim 1$, the primary CMB 
to $z\sim 1089$ whereas CMB lensing
probes $1\lesssim z\lesssim 3$.  Given that spatial curvature affects distances 
and growth out
to high redshift, CMB lensing is ideally suited to breaking the degeneracy between
curvature and the dark energy. It has the additional advantage of being nearly
entirely in the linear regime and a lensing test of curvature where the source
distance can be considered fixed.

Furthermore, this degeneracy breaking requires only already planned ground-based
CMB polarization power spectrum measurements.  
We have demonstrated that even if the SNAP and Planck surveys are limited
only by statistical errors, 
a ground based survey like SPTpol will be sufficient to
extract the full information: $\sigma(w_p) = 0.02$, $\sigma(w_a)=0.2$ and $\sigma(\Omega_K)=0.0034$; with some accounting for SNAP systematic 
errors these degrade to $0.025$, $0.3$, and $0.0035$.

There are two critical assumptions that make this possible.  Firstly
that the ground based CMB survey will be able to remove
foregrounds and systematics 
at a level sufficient to enable few percent level measurements of the lensing
$B$-mode polarization power.
Secondly, we assume that the neutrino masses are fixed by oscillation measurements
and a theoretical assumption about the neutrino mass hierarchy. 
This assumption will be tested by next generation laboratory experiments.
In the more general context, the sum of the neutrino masses must be externally
determined to $0.1$eV or better.

The lensing observables approach we have taken here can be easily extended to
consider different combinations of probes or alternate explanations of the
accelerated expansion.  
 Furthermore we have only considered the simplest description of the time-dependent dark
energy density, in terms of parameters $w_0$ and $w_a$. More
ambitious descriptions of the dark energy sector or more
exotic theoretical models with high redshift deviations may be even further assisted by CMB lensing.

\acknowledgements {\it Acknowledgments}: We thank Manoj Kaplinghat and
 Kathryn Miknaitis for useful
conversations.  WH and KMS were supported by the KICP
through the grant NSF PHY-0114422. WH was additionally
supported by U.S.~DOE contract DE-FG02-90ER-40560 and the David and
Lucile Packard Foundation. DH was supported by the NSF Astronomy and
Astrophysics Postdoctoral Fellowship under Grant No.\ 0401066.


\end{document}